%% file: main.tex
\documentclass[sigconf, nonacm]{acmart}

\newcommand\vldbdoi{XX.XX/XXX.XX}
\newcommand\vldbpages{XXX-XXX}
\newcommand\vldbvolume{14}
\newcommand\vldbissue{1}
\newcommand\vldbyear{2020}
\newcommand\vldbauthors{\authors}
\newcommand\vldbtitle{\shorttitle} 
\newcommand\vldbavailabilityurl{https://github.com/ZhaoFuheng/rocksdb}
\newcommand\vldbpagestyle{plain}

\usepackage{xspace}
\newcommand{\policy}{Garnering\xspace}
\newcommand{\policyR}{Garnering\xspace}

\newcommand{\smartparagraph}[1]{\noindent{\bf #1}\ }
\usepackage{graphicx}
\usepackage{subfig}
\usepackage{adjustbox}

\begin{document}
\title{Autumn: A Scalable Read Optimized LSM-tree based Key-Value Stores with Fast Point and Range Read Speed}

\author{Fuheng Zhao}
\affiliation{%
  \institution{UC Santa Barbara}
}
\email{fuheng\_zhao@ucsb.edu}

\author{Zach Miller}
\affiliation{%
  \institution{UC Santa Barbara}
}
\email{zmiller@ucsb.edu}

\author{Leron Reznikov}
\affiliation{%
  \institution{UC Santa Barbara}
}
\email{lreznikov@ucsb.edu}

\author{Divyakant Agrawal}
\affiliation{%
  \institution{UC Santa Barbara}
}
\email{agrawal@cs.ucsb.edu}

\author{Amr El Abbadi}
\affiliation{%
  \institution{UC Santa Barbara}
}
\email{amr@cs.ucsb.edu}







\begin{abstract}
Log Structured Merge Trees (LSM-tree) based key-value stores are widely used in many storage systems to support a variety of operations such as updates, point reads, and range reads. Traditionally, the merge policy of LSM-trees organizes data into multiple levels of exponentially increasing capacity to support high-speed writes. However, we contend that the traditional merge policies are not optimized for reads. In this work, we present Autumn, a scalable and read-optimized LSM-tree based key-value store with near-optimal worst-case point and range read costs. The key idea in improving read performance is to dynamically adjust the capacity ratio between two adjacent levels as more data are stored. As a result, lower levels gradually increase their capacities and merge more often. In particular, point and range read cost improves from the previous best known $O(logN)$ complexity to $O(\sqrt{logN})$ in Autumn by applying the novel \textit{Garnering} merge policy. While the \textit{Garnering} merge policy optimizes for both point reads and range reads, it maintains high performance for writes by inherently prioritizing the merges in the lower levels, as Garnering schedules more merges for the lower levels. We implemented Autumn on top of RocksDB and LevelDB and experimentally show the gain in performance for real-world workloads.
\end{abstract}

\maketitle

\pagestyle{\vldbpagestyle}
\begingroup\small\noindent\raggedright\textbf{PVLDB Reference Format:}\\
\vldbauthors. \vldbtitle. PVLDB, \vldbvolume(\vldbissue): \vldbpages, \vldbyear.\\
\href{https://doi.org/\vldbdoi}{doi:\vldbdoi}
\endgroup
\begingroup
\renewcommand\thefootnote{}\footnote{\noindent
This work is licensed under the Creative Commons BY-NC-ND 4.0 International License. Visit \url{https://creativecommons.org/licenses/by-nc-nd/4.0/} to view a copy of this license. For any use beyond those covered by this license, obtain permission by emailing \href{mailto:info@vldb.org}{info@vldb.org}. Copyright is held by the owner/author(s). Publication rights licensed to the VLDB Endowment. \\
\raggedright Proceedings of the VLDB Endowment, Vol. \vldbvolume, No. \vldbissue\ %
ISSN 2150-8097. \\
\href{https://doi.org/\vldbdoi}{doi:\vldbdoi} \\
}\addtocounter{footnote}{-1}\endgroup

\ifdefempty{\vldbavailabilityurl}{}{
\vspace{.3cm}
\begingroup\small\noindent\raggedright\textbf{PVLDB Artifact Availability:}\\
The source code, data, and/or other artifacts have been made available at \url{\vldbavailabilityurl}.
\endgroup
}

\input{introduction}
\input{background}
\input{Autumn}
\input{experiments}
\input{relatedWorks}
\input{conclusion}



\bibliographystyle{ACM-Reference-Format}
\bibliography{main-bib}

\end{document}

%% file: introduction.tex
\section{Introduction}

The Log Structured Merge Tree (LSM-tree) is a popular structure for the storage layer in modern key-value stores and large scale data systems. By design, LSM-tree offers consistently high write throughput by applying out-of-place updates. It buffers all updates in main memory, stores the updates into a write-ahead log (WAL), and flushes all the buffered updates into disk as a sorted run when the buffer reaches its capacity. This process transforms random disk writes into sequential disk writes and significantly improves write performance. As a result, LSM-trees have become the cornerstone for industrial key-value store engines~\cite{chatterjee2021cosine}, including BigTable~\cite{chang2008bigtable}, Dynamo~\cite{decandia2007dynamo}, HBase~\cite{george2011hbase}, Cassandra~\cite{lakshman2010cassandra}, RocksDB~\cite{dong2021rocksdb}, and LevelDB~\cite{levelDB}. In addition, LSM-tree-based storage is commonly found in graph processing~\cite{sharma2016dragon}, stream processing~\cite{chen2016realtime}, Online Transactional Processing~\cite{matsunobu2020myrocks}, and for supporting real-time fresh analytics in Hybrid Transactional-Analytical Processing (HTAP). Popular HTAP architectures, such as ByteHTAP~\cite{chen2022bytehtap} and TiDB~\cite{huang2020tidb}, use two separate storage engines to process On-Line Transactional Processing (OLTP) and On-Line Analytical Processing (OLAP) workloads. In ByteHTAP~\cite{chen2022bytehtap}, DMLs, and DDLs, simple queries are sent to its OLTP engine. Similarly, in TiDB~\cite{huang2020tidb}, small range queries and point queries, can be sent to the TiKV storage engine~\cite{TiKV}, which is an LSM-tree based key-value storage engine with row-oriented layout to handle such workloads.


An LSM-tree organizes data into multiple levels with increasing capacities and the heighest level stores the majority of the data. The merge policy in an LSM-tree explicitly defines when data should be moved across two adjacent levels. Many existing works have focused on minimizing space amplification~\cite{dong2017optimizing}, reducing the write amplification~\cite{dayan2019log}, and optimizing for point queries~\cite{dayan2018dostoevsky} by introducing new merge policies. The merge policy in LSM-trees implicitly defines the trade-off among the storage efficiency (i.e., with out-of-place updates obsolete data entries may be stored across different levels), I/O costs of updates, and the I/O costs of reads. Mainstream LSM-tree storage engines often use either \textit{Tiering} or \textit{Leveling} merge policies~\cite{levelDB, TiKV, dong2021rocksdb, lakshman2010cassandra}. On one hand, the Tiering policy gradually sort-merges runs in a level and it is write optimized. On the other hand, the Leveling policy actively sort-merges runs in a level and is read optimized. However, we contend that existing designs are sub optimal for range reads, and hence, we propose Autumn with \policy policy to provide more superior read performances than existing solutions.


In addition to the merge policies, existing solutions for optimizing read queries primarily use better caching strategy~\cite{teng2018low, wu2020ac} and probabilistic data sketches/filters~\cite{zhang2018elasticbf, luo2020rosetta, zhang2018surf, knorr2022proteus, zhao2023panakos}, in which filters are stored in main memory to reduce disk I/Os by skipping sorted runs with no records inside the search key range. These approaches are orthogonal to our work. However, range filters do not improve range query performance when each sorted run may contain some portions of the range query results and hence all runs need to be accessed. Also, range filters do not offer both good range filter query functionality and the desired low false positive rates on point filter queries at the same time. With a fixed filter memory budget, range filters have worse performance on point filter queries compared to the traditional point filter~\cite{bloom1970space}. As a result, range filters cannot directly replace point filters in many LSM-tree storage engines when their workloads contain numerous point queries.

Unlike prior works, Autumn focuses on improving both range queries and point queries at the same time with strong worst-case theoretical complexity guarantees while keeping updates scalable. Autumn uses the new \textit{Garnering} policy to optimize range queries and achieve the best range query complexity among existing merge policies. Autumn also uses the bloom filter optimizations inspired from Monkey~\cite{dayan2017monkey} to realize state-of-the-art point query performance. By offering low latencies for point and short range queries and scalable update costs, Autumn is well suited for both OLTP and HTAP workloads. Specifically, we propose Autumn, an LSM-based storage engine that is designed for optimizing reads in key-value stores with the following goals:
\begin{itemize}
    \item \textbf{Performant Range Query}. Range reads are important for analytics and OLTP workloads. Autumn focuses on fundamentally improving the worst-case complexities for range reads by minimizing the number of disk access.

   \item \textbf{Fast Point Query}. Point queries are very common. When no probabilistic point filters are used to skip runs in a point query, Autumn should also improve the worst-case point query performance. In addition, probabilistic data structures play a significant role in improving point query performance. Like prior works, Autumn leverages and optimizes point filters stored in main memory to increase the performance of point query.
   
   \item \textbf{Scalability}. As more and more key-value pairs are stored, Autumn should scale up and provide good space amplifications and update performance, while excel in read operations.
 \end{itemize}

In this paper, we propose the novel Autumn key-value store, which redesigns the traditional LSM-tree data organization with the \policy policy and focuses on optimizing reads while keeping updates scalable. In summary, the main contributions in this paper are:
\begin{enumerate}
  \item We introduce Autumn with the new \policy merge policy to redesign the data organization in LSM-trees. Among all existing merge policies, \policy achieves the best range query cost, enhances point query performance with no bloom filters, and maintains the state-of-the-art point query cost when bloom filters are used.

  \item We apply the disk I/O cost model to provide thorough theoretical analysis on the complexities of range query, point query, write amplification, and space amplification.

  \item We find that Autumn's write and update performance are further optimized by keeping a constant number of sorted runs in the first level of LSM-tree to avoid all merges at the first level without affecting the read complexity. This approach is in fact implemented in LevelDB and RocksDB, and can significantly reduce write amplification in Autumn. Since Autumn gradually increases the capacity of lower levels as more data entries are stored, the compactions derived from the first level consist of a large fraction of the total write amplifications.

  \item We conduct thorough experiments to demonstrate Autumn's high performance in reads and scalability in writes and reads with RocksDB and LevelDB using the default benchmark~\cite{levelDB} and Yahoo!’s YCSB benchmark to emulate real-world workloads~\cite{cooper2010benchmarking}.

\end{enumerate}

The paper is organized as follows: Section~\ref{sec:background} discusses the background information of LSM-trees and gives an overview of existing merge policies and their characteristics. Section~\ref{sec:autumn} introduces our proposed scalable and read optimized key-value store Autumn with the novel \policy merge policy and we present detailed analysis of the complexity bounds for Autumn in space amplification, reads, and updates costs. Section~\ref{sec:eval} presents the experimental results of an evaluation conducted using the YCSB benchmark. Section~\ref{sec:relatedWorks} discusses related works that are orthogonal to Autumn ranging from range filters to new hardware. Finally, Section~\ref{sec:conclusion} summarizes our contributions and concludes this work.

%% file: background.tex
\section{LSM-tree Background} \label{sec:background}

In this section, we present background information on LSM-trees. O'Neil et al.~\cite{o1996log} introduced the LSM-tree in 1996 to reduce random disk access during updates in database index structures by applying out-of-place updates. Prior to LSM-trees, out-of-place updates were successfully supported in Differential Files, Postgres project, and Log-Structured File System~\cite{rosenblum1992design, severance1976differential,stonebraker1987design} where large amounts of data were collected before writing the data to a file in a single large I/O. However, these approaches of sequentially writing data into an append-only log hurt query performance, and there is no principled cost analysis of trade-offs among writes, reads, and space utilization. In Table~\ref{tab:Table of Symbols}, we list the important symbols used to analyze LSM-trees in this paper. 

\begin{figure*}[tbph]
\centering
\includegraphics[width=\textwidth]{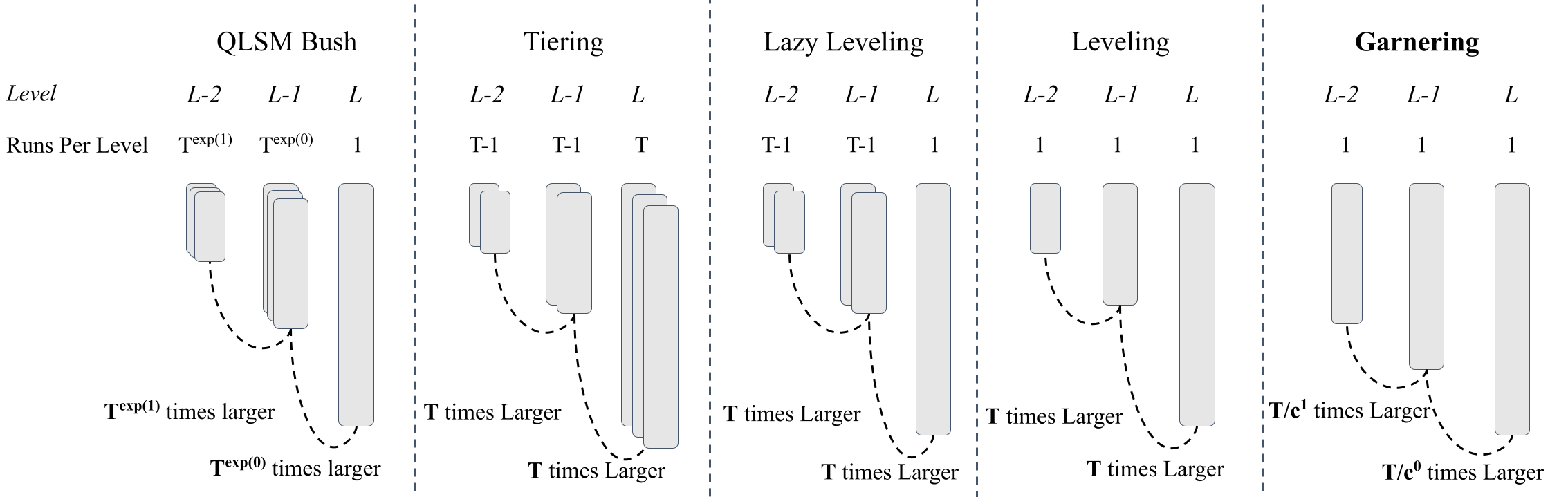}
\caption{An illustration of the last three largest levels with QLSM-Bush, Tiering, Lazy-Leveling, Leveling, and our proposed \policy. The left is more write optimized. The right is more read optimized.}
\label{fig:MergePolicies}
\end{figure*}

LSM-trees provide a systematic methodology to analyze the trade-offs among writes, reads and space utilization. An LSM-tree physically stores data into $L$ levels where level 0 resides in memory, and the other levels reside in persistent storage. The capacity of each level (the amount of data that can be stored at each level) increases exponentially. The capacity at level $i$ is $T$ times larger than the capacity at level $i-1$, as shown in Equation~\ref{eq:ori-ratio}.
\begin{equation} \label{eq:ori-ratio}
    C_{i}/C_{i-1} = T
\end{equation}

In the original design~\cite{o1996log}, each level is structured as a $B^{+}$-Tree, and when level $i$ reaches its capacity, the rolling merge policy moves a range of $B^{+}$-Tree leaf pages from level $i$ to level $i+1$. In modern LSM-trees, new updates are always stored in the level 0 memory table (\textit{memtable}), which is often implemented using skiplists to support both point and range reads. During updates, if a data entry with the same key is already stored, the new value will replace the old value, and if the key is deleted, the LSM-tree associates a tombstone with the key. When the memtable is full, it becomes immutable, and the data stored in memtable will be flushed as a sorted \textbf{run} stored in the static sorted table (SST) at level 1, which resides on disk. A run logically is a sorted array based on the key, and physically it consists of one or more SST files. When the data stored at level $i$, $N_{i}$, reaches the capacity of level $i$, $C_{i}$, data files at level $i$ will be merged into level $i+1$ with a larger capacity. During the merge, entries with duplicate keys will store the newer value, and the older value will be discarded.

\begin{table}[htbp]
    \begin{center}
    \begin{tabular}{r l }
    \toprule
    $N$ & Data size\\
    $B$ & Memtable size\\
    $L$ & Number of levels in LSM-tree\\
    $N_{i}$ & Data size at level $i$\\
    $C_{i}$ & Capacity at level $i$\\
    $T$ & Size ratio in merge polices\\
    $c$ & The scaling ratio used in \textit{Garnering} policy.\\
    $M_{filter}$ & Memory budget for Bloom Filter\\
    \bottomrule
    \end{tabular}
    \end{center}
    \caption{Table of symbols}
    \label{tab:Table of Symbols}
\end{table}

\subsection{Concurrency Control and Recovery}
A strict requirement for usability is to ensure correctness of the storage engine with concurrent requests (reads and updates) and when the machine crashes. LSM-trees typically employ a multi-version concurrency control protocol instead of locking to avoid contention~\cite{luo2020lsm}. After a  flush operation or a compaction, the files involved will have a new version, and old files with obsolete versions can be garbage-collected. A point and range query will first retrieve a set of files with the newest version and use this set of files to return the desired and consistent result. Since all updates are first buffered in main memory, the write-ahead log (WAL) is naturally used in LSM-trees to ensure durability in case of a machine failure. Also, there is a metadata log to store the state of the LSM-tree's structure. During the recovery phase, the key-value store will then redo all committed transactions from the transaction log, and it can use the metadata log to recover all the files stored on the disk.

\subsection{LSM-tree Amplification}
\textbf{Write Amplification} Write Amplification represents the write cost, and it measures the amortized number of disk writes for adding a data entry into an LSM-tree. An LSM-tree organizes data into multiple levels. Smaller levels collect data and write the collection of data into larger levels in a single sizeable sequential I/O. As a result, data entries are written to storage multiple times as the LSM-tree grows. Repeated writes can negatively affect the lifespan of modern solid-state storage devices (SSDs)~\cite{min2012sfs, lee2015f2fs}. Hence, high write amplification can wear out the device. As a result, researchers have studied and successfully developed different mechanisms such as separating key and value~\cite{lu2017wisckey} and introducing a non-volatile memory layer between main memory and disk~\cite{im2020pink} to reduce the amount of write amplifications. To avoid write stalls caused by long merge operations at larger levels, LSM-tree based key-value stores, such as LevelDB~\cite{levelDB} and RocksDB~\cite{dong2021rocksdb}, perform partial merges, which move the SST file (2-64 MB) between two adjacent levels instead of an entire sorted run. This mechanism tries to distribute the merging overheads over time and can negatively impact the worst-case update I/O cost of merging~\cite{dayan2022spooky}. With full merge, two runs in adjacent levels are merged entirely at once and full merges are used as default in Cassandra~\cite{lakshman2010cassandra} and HBase~\cite{george2011hbase}. Throughout this paper, we discuss the complexity of merge operations as having the granularity of runs.

\textbf{Point Query Amplification} A point query finds the most recent version of a data entry by traversing all levels in the LSM-tree, and since the most up to date information is stored in the lower levels, the query will terminate once it finds an entry with the matching key. A common practice to speed up point queries is to associate each sorted run with a \textbf{Bloom Filter}~\cite{bloom1970space}, which is stored in main memory. A bloom filter is a space-efficient data sketch that answers membership queries. In addition, it guarantees no false negatives answers, i.e., a data entry is not stored in the data set if the membership query returns false, and it has small tunable false positive rates (FPR), i.e., a data entry may not be stored in the data set while the membership query returns true. The FPR depends on the number of bits used per data entry, and using 10 bits per data entry sets the FPR around 1$\%$~\cite{tarkoma2011theory}, as expressed in Equation~\ref{eq:bloom FPR} where $M_{filter}$ is the filter memory size and $N$ is the number of data entries.
\begin{equation} ~\label{eq:bloom FPR}
    FPR = e^{-ln(2)^{2}M_{filter}/N}
\end{equation}

\textbf{Range Query Amplifications} A range query finds the most recent version of all data entries from a starting target key until satisfying a specific condition, such as the iterator reaching the end key boundary or the keys returned from the iterator reach a desired number. Unlike point queries that can return once the target key is found, range queries need to examine all levels to ensure that no keys within the target key range are missing. A merging iterator will collect the newest version of the relevant data entries in the target key range from iterators associated with each sorted run, and each sorted run's iterator is responsible for reading the data entries until the ending criterion is met. A common practice to speed up the iterator seek time is to store \textbf{Fence Pointers} in main memory. Fence pointers store the first key of every block of every run in main memory such that a relevant key range for a run can be located with one disk I/O.

\textbf{Space Amplification} Since the LSM-tree uses out-of-place updates, obsolete data entries are stored on disk and amplify the total storage. Hence, an LSM-tree requires larger storage space than the logical data sizes. Space Amplification is represented as the ratio between the physical storage size and the logical storage size. Traditionally, space amplification raises little concern due to the affordability of storage devices. In the era of cloud-native databases, space amplification can increase the operational cost. To present the full metrics in the design space, we also aim to keep space amplification small.

\begin{table*}[tbph]

    \begin{center}
     \begin{tabular}{l | l | l | l | l| l} 
     \hline
     Policy &  Range Query & Point Query w/o Filter & Point Query w/ Filter & Write &  Space\\  
     \hline\hline

     QLSM Bush & $O(\sqrt{T \cdot \frac{N}{B}})$ & $O(\sqrt{T \cdot \frac{N}{B}})$ & $O(T \cdot e^{-\frac{M_{filter}}{N}})$ & $O(1 + log_2(log_T(\frac{N}{B})))$ & $O(1)$\\
     \hline

     Tiering & $O(T \cdot log_{T}(\frac{N}{B}))$ & $O(T \cdot log_{T}(\frac{N}{B}))$ &  $O(T \cdot e^{-\frac{M_{filter}}{N}})$ & $O(\frac{log_{T}(\frac{N}{B})}{B})$ & $O(T)$\\
     \hline

     Lazy-Leveling & $O(1 + T\cdot log_{T}(\frac{N}{B}))$ & $O(1 + T\cdot log_{T}(\frac{N}{B}))$ & $O(e^{-\frac{M_{filter}}{N}})$ & $O(T + \frac{log_{T}(\frac{N}{B})}{B})$ & $O(\frac{T+1}{T})$\\
     \hline

     Leveling & $O(log_{T}(\frac{N}{B}))$ & $O(log_{T}(\frac{N}{B}))$ & $O(e^{-\frac{M_{filter}}{N}})$ & $O(T \cdot \frac{log_{T}(\frac{N}{B})}{B})$ & $O(\frac{T+1}{T})$ \\ 
     \hline
     
     \policyR (Ours) & $O(\sqrt{-log_{c}(\frac{N}{B \cdot T})})$, $c<1$) & $O(\sqrt{-log_{c}(\frac{N}{B \cdot T})})$ & $O(e^{-\frac{M_{filter}}{N}})$ & $O(T/c^{\sqrt{-log_{c}(\frac{N}{B \cdot T})}})$ & $O(\frac{T+1}{T})$ \\
     \hline

     
    \end{tabular}
    \end{center}
\caption{Complexity Analysis among existing policies and \policy.}
\label{tab:policy complexity}
\end{table*}

\subsection{Merge Policies}
The merge policies determine when data should be merged between two adjacent levels in LSM-tree. Moreover, the merge policies have a direct impact on point query cost, range query cost, write cost, and space amplification. There is an inherent trade-off among these metrics.

\subsubsection{Leveling and Tiering}
Figure~\ref{fig:MergePolicies} illustrates the state-of-the art merge polices. Both Leveling and Tiering merge policies exponentially increase level capacities from small levels to large levels in which two adjacent levels' capacity differs by a constant factor of $T$ (a tunable knob) and $T>1$.

\smartparagraph{Number of levels.} In traditional policies: Since the capacity of the level grows exponentially, the last level has a capacity of $BT^{L}$, where $B$ is the memtable size and L is the total number of levels. The sum of data stored at all levels besides the last level is approximately equal to the data stored at the last level, and the amount of data stored at the last level is $N\frac{T-1}{T}$, where $N$ is the size of the data stored in the LSM Tree. Hence, by solving the equation $BT^{L} = N\frac{T-1}{T}$ for $L$, the number of levels can be calculated as $O(log_{T}(N/B))$, as shown in~\ref{eq: classicNumLevels}.
\begin{equation} \label{eq: classicNumLevels}
    L = O(log_{T}(\frac{N}{B} \cdot \frac{T-1}{T}))
\end{equation}

In the \textbf{Leveling} merge policy~\cite{levelDB, dong2021rocksdb}, there is at most one sorted run for each level. Merge is triggered immediately as the new run arrives. When level $l$ reaches its capacity $C_{l}$, then level $l$'s sorted run will go through a compaction such that level $l$'s run will be sort-merged with level $l+1$'s run. As a result, this design requires data to be merged on average $O(T)$ times within each level before the level's capacity is reached. The write amplification can be derived as $O(T \cdot L)$.  For range queries, all runs need to be accessed. Since there is one run per level, the range query cost is $O(L)$ disk I/O. The worst-case point read query cost happens when the target key is not stored. Therefore, in the worst case all runs need be accessed and the cost becomes $O(L)$ disk I/O. With a bloom filter attached to each run, the number of disk I/O cost is the sum of all false positive rates across all bloom filters. Assuming all bloom filters use the same number of bits per data entry, then the worst-case point read query cost becomes $O(L \cdot e^{-M_{filter}/N})$.

In the \textbf{Tiering} merge policy~\cite{lakshman2010cassandra}, each level can store at most $T$ sorted runs. When level $l$ becomes full, $T$ runs will be sort-merged and then stored as a new sorted run at level $l+1$. As a result, this design only merges one time at each level before the level's capacity is reached, and the write amplification is $O(L)$. All runs need to be accessed for range query, and hence the cost is $O(T \cdot L)$ disk I/O. For the worst-case point read query cost, the point read cost is $O(T \cdot L)$ disk I/O assuming no bloom filters are used. If all bloom filters use the same amount of bits, the worst-case point read cost becomes $O(T \cdot L \cdot e^{-M_{filter}/N})$, since there are $T$ bloom filters for each level. On the one hand, the Leveling policy actively sort-merges runs to reduce the total number of runs and is more read optimized. On the other hand, the Tiering policy passively sort-merges runs to reduce the update cost.

\smartparagraph{Bloom Filter Optimization.} The LSM-tree implementations in industry have traditionally set the FPR uniformly across all levels, setting $O(ln(1/\epsilon))$ bits per data entry at all levels. As a result, larger sorted runs need exponentially more filter memory budget to save one disk I/O, as Fence Pointers can locate a given data block. For point queries, the worst-case disk I/O is the sum of all FPR from all runs in which the target query key is not stored in the database. Monkey~\cite{dayan2017monkey} proposes a better approach by decreasing the FPR rates at lower levels. The central intuition is that the cost of accessing the disk in each level is the same. Thus minimizing the FPRs at lower levels is much more memory budget efficient as they have exponentially fewer data entries and require smaller filter memories to achieve low FPRs compared to larger levels. As a result, this design improves the worst-case point read query complexity by a factor of $L$. With the bloom filter optimization, the worst-case point read query cost becomes $O(e^{-\frac{M_{filter}}{N}})$ and $O(T \cdot e^{-\frac{M_{filter}}{N}})$ for Leveling and Tiering respectively, which are independent of the number of levels and only depend on the filter memory budget ($M_{filter}$).

\subsubsection{Other Merge Policies}
A series of works~\cite {dayan2018dostoevsky,dayan2019log} focused on proposing new merge policies to achieve better point query cost, while maintaining scalable write performance. However, these policies are not fully read optimized as they only improve point reads and not range reads. In particular, Lazy-Leveling~\cite{dayan2018dostoevsky} takes inspiration from both Leveling and Tiering and keeps the last level as one sorted run and all other levels as $T$ sorted runs. Hence, write amplification becomes $O(T + \frac{log_{T}(N/B)}{B})$ and the range read cost is $O(1 + T \cdot log_{T}(N/B))$, which has worse range read performance than Leveling. QLSM Bush~\cite{dayan2019log} achieves the best write amplification $O(1 + log_2(log_T(N/B)))$ while sacrificing range reads to $O(\sqrt{TN/B})$. As shown in Figure~\ref{fig:MergePolicies}, QLSM Bush organizes data with doubly exponential more runs at lower levels to achieve very low write amplifications. As a result, range read cost becomes $O(\sqrt{T \cdot N/B})$ in which the smallest level contains roughly $O(\sqrt{T \cdot N/B})$ sorted runs. With bloom filter optimizations, Lazy-Leveling and QLSM Bush reduce the point query cost to $O(e^{-\frac{M_{filter}}{N}})$ and $O(T \cdot e^{-\frac{M_{filter}}{N}})$ respectively. It is clear that the existing merge policies are not optimized in terms of range query performance. In contrast, Autumn focuses on optimizing for both point and range queries. By introducing the \textbf{\policy} merge policy, which uses an additional scaling factor $c$ between levels, Autumn simultaneously matches the best point query cost with and without bloom filters and provides the best worst-case range read performance, while maintaining scalable space amplification and write amplification. In the next section, we describe the new \policy policy and the architecture of Autumn in more detail. Table~\ref{tab:policy complexity} shows the trade-offs among the different policies.


%% file: Autumn.tex
\section{Autumn} \label{sec:autumn}


In this section, we introduce Autumn a new design for LSM-trees with \policy Merge Policy that optimizes for both point and range queries and offers good performance in writes and space amplifications.

\subsection{A New Design}

In Autumn, we rethink the data organization of LSM-trees and propose a new merge policy, \policy. We aim to fundamentally improve the worst-cast point and range read complexity while keeping high write performance. When bloom filters are applied, prior studies have improved point reads efficiency by tuning the bloom filter memory budget at each level to minimize the worst-case false positive rate~\cite{dayan2017monkey}. In addition, a large number of works have aimed to reduce write amplification~\cite{dayan2019log, yao2020matrixkv, lu2017wisckey}. However, optimizing the cost for point and range reads remains to be a challenge. Our design is focused on optimizing for both range queries, and also for point queries, with and without bloom filters, while offering high throughput for updates. Our main intuition is that to improve range query efficiency, we need to decrease the number of levels more aggressively as the database size grows. In addition, to make sure the design is scalable, write amplification needs to have sub-linear complexity as a function of $N$.

\smartparagraph{Garnering High-Level Design.} Instead of exponentially incrementing each level's capacity with the constant factor $T$, \policy flattens the LSM-tree by only fixing the capacity ratio difference between the last level $L$ and level $L-1$ to a constant factor $T$ and increasing the ratio difference between lower levels with a scaling factor $c$ where $c$ is less than 1, as shown in Equation~\ref{eq:new ratio diff}. The reason why we set $c<1$ will be apparent when analyzing write amplification in which the sum of write amplifications among different levels form a geometric sequence such that the compaction cost at the small levels dominate. In essence, the capacity ratio difference between two adjacent layers dynamically changes when more data are stored.
\begin{equation} \label{eq:new ratio diff}
    C_{i}/C_{i-1} = T / c^{L-i}
\end{equation}
The capacity at level $i$ follows Equation~\ref{eq:new capacity}. Once data stored in level $i$ reaches its capacity, files from level $i$ will be merged into level $i+1$. As a result, when the data storage grows, the lower levels will have a larger capacity compared to the traditional data organization.
\begin{equation} \label{eq:new capacity}
    C_{i} = T^{i} / c^{(2L-1-i)i/2} \cdot B
\end{equation}

\smartparagraph{Delayed Last Level Compaction.} Traditionally, the capacity of each level does not change as more data are stored and has no dependency on the number of levels. However, in our new \policy policy, we gradually increase the capacity of lower levels as more data are stored. As a result, the capacity of each level depends on the total number of levels. Usually, when the last level $l$ is full, a new level $l+1$ should be created. Since in \policy, the capacities of levels increase when a new level is created, the compaction of level $l$ is not necessary, as the new capacity at level $l$ with respect to the total number of levels $l+1$ is strictly larger than the data stored, i.e., $C_{l}>N_{l}$ after a new level is created. Therefore, when the last level is full and a new level needs to be created, we only increment the total number of levels by one and then delay the compaction to the next cycle. In this approach, we avoid performing unnecessary last level compactions and improve write performance.

\smartparagraph{Space Amplification.} Recall that space amplification is defined as the ratio between the physical storage size with the logical storage size, and smaller space amplification translates to better efficiency in term of disk storage. Since each level only contains one sorted run, in the worst case scenario, all entries stored in level 1 to level $L-1$ are all updates to existing data entries stored in level $L$. To analyze the worst case complexity bound of space amplification, we first observe that the sum of all the data stored in level 1 to level $L-1$ is dominated by the capacity at level $L-1$, i.e., $C_{L-1}$. We also know that the last level $L$ is $T$ times larger than level $L-1$. Hence, in the worst case scenario, $O(1/T)$ of entries in the last level are obsolete. Hence, the space amplification is $O(\frac{T+1}{T})$. 

In the analysis, we assume all levels are filled to their capacity. In fact, when the last level is not filled to its capacity, the space amplification may be larger. As a result, to minimize space amplification, LSM-tree based key-value stores often dynamically adjust the capacity of lower levels based on the amount of data stored in the last level (forcing the last level to be full) such that the data stored in level 1 to level $L-1$ is a small fraction of the data stored in the last level~\cite{dayan2022spooky,dong2017optimizing}.

\smartparagraph{Number of Levels in \policy.} Since the capacity is modeled in Equation~\ref{eq:new ratio diff}, the capacity at the last level is $T^{L}/c^{(L(L-1)/2)} \cdot B$. Since the capacity of the last level $L$ is $T$ times larger than the capacity of level $L-1$, the approximate data size at the last level is $N \cdot \frac{T-1}{T}$. Hence, by solving the equation $T^{L}/c^{(L(L-1)/2)} \cdot B = N \cdot \frac{T-1}{T}$ for $L$, we obtain the number of levels as $O(\sqrt{-log_{c}(\frac{N}{B \cdot T})})$, as shown in Equation~\ref{eq:newNumLevel}. Since $c$ is less than 1, $log_{c}$ of a positive real number yields a negative value and hence there is a negative sign inside the square root.~\footnote{Note, $-log_{c}(x) = ln(x)/-ln(c) = ln(x)/ln(1/c) = log_{1/c}(x)$.}
\begin{equation} \label{eq:newNumLevel}
    L = O(\sqrt{-log_{c}(\frac{N}{B} \cdot \frac{T-1}{T})} - log_{1/c}(T))\\
\end{equation}
\begin{equation*} \label{eq:autumnlevels}
    L = O(\sqrt{-log_{c}(\frac{N}{B\cdot T})})
\end{equation*}

\smartparagraph{Point Query.} In LSM-Trees, the worst-case point read cost is $O(L)$ with zero look-ups. When the target key is not stored in the storage engine, all levels must be examined before returning the answer (key is not found). Since the number of levels is $O(\sqrt{logN})$ in Autumn, as shown in Equation~\ref{eq:autumnlevels}, Autumn with \policy enhances the performance of point query to $O(\sqrt{logN})$ complexity. Moreover, bloom filters may speed up point query performance, allowing runs without the target key to be skipped. Disk I/0s are saved at some moderate main memory space cost in which the filters require about 2 to 10 bits per data entry.

\smartparagraph{Bloom Filter Optimization.} We showcase how \policy achieves the same worst case point query complexity as the complexity in Leveling by fine-tuning the bloom filters memory budget across different levels. The method to derive the optimal bloom filter memory allocations is inspired by Monkey~\cite{dayan2017monkey}. 

The expected number of probes wasted by a zero-result lookup is the summation of FPRs from all Bloom filters. In \policy,  each level contains one sorted run. The point read cost $R$ can be calculated using Equation~\ref{eq:read cost}, where it sums all the FPRs from level 1 to level $L$.
\begin{equation} \label{eq:read cost}
R = \sum_{i=1}^{L}\ p_{i} 
\end{equation}
where $p_{i}$ is the false positive rate for the bloom filter at level $i$ and $p_{i}$ is between 0 and 1.

Then, we can calculate the total filter memory by summing each level's filter memory. Based on Equation~\ref{eq:bloom FPR}, we can rearrange the terms such that the filter memory needed for level $i$ is determined by $p_{i}$ and $N_{i}$. The necessary filter memory for level $i$ is $M_{i} = -N_{i} \frac{ln(p_{i})}{ln(2)^2}$. Each level's filter memory size is proportional to the the level's capacity and the corresponding false positive rate, and hence we can apply Equation~\ref{eq:new ratio diff} to calculate the ratio factor between lower levels to the highest level. As a result, the total filter memory size is shown in Equation~\ref{eq:new total filter memory}.

\begin{equation} \label{eq:new total filter memory}
    M_{filter} = - \frac{N}{ln(2)^2} \cdot \frac{T-1}{T} \sum_{i=1}^{L} ln(p_{i}) \frac{c^{(L-i)(L-i-1)/2}}{T^{L-i}}
\end{equation}

Then, we can formulate the optimal bloom filters memory allocations into an optimization problem. The goal is to minimize the read cost $R$, which is the sum of FPRs across all levels as shown in Equation~\ref{eq:read cost}, given the total memory budget $M_{filter}$, as shown in Equation~\ref{eq:new total filter memory}. Given the equality constraint, we apply the commonly used Lagrange approach~\cite{bertsekas2014constrained} to minimize the cost. By taking the partial derivative of the Lagrangian expression with respect to $\{p_1, p_{2}, ..., p_{L}\}$ and setting the derivatives to zero, we derive the best FPRs for each level as a function of $p_{L}$, as shown in Equation~\ref{eq:p_i(p_L)}.

\begin{equation} \label{eq:p_i(p_L)}
    p_{L-i} = p_{L} \cdot \frac{c^{i(i-1)/2}}{T^i} 
\end{equation}

Each FPRs in the read cost $R$ can be rewritten by plugging in Equation~\ref{eq:p_i(p_L)}, such that 
\begin{equation*}
    R = p_{L} + p_{L} \cdot \frac{c^{0}}{T} +  p_{L} \cdot \frac{c^{0+1}}{T^2} +  p_{L} \cdot \frac{c^{0+1+2}}{T^3} + ...
\end{equation*}
By closely examining the new expression above for the read cost, we observe that if the numerator is always 1, then this is a geometric series. Here, the numerator is some power of $c$. Since $c$ is a small constant less than 1, this series actually converges much faster than a geometric sequence and as L grows, the read cost quickly converges to $R = O(p_{L})$.

Therefore, given a desired read cost, we can derive the optimal false positive rates as shown in Equation~\ref{eq:new bloom allocations}.
\begin{equation} \label{eq:new bloom allocations}
    p_{i} = O(R \cdot \frac{c^{(L-i)(L-i-1)}/2}{T^{L-i}})
\end{equation}

\smartparagraph{Filter Memory Budget.} Assuming the filter memory budget is fixed, then as the amount of data grows, the FPRs increase since a filter's FPR has a linear dependence on $M_{filter}/N$. Eventually, level $L$'s false positive rate will converge to one. Therefore, it is useful to analyze the ratio of bits per data entry, $M_{filter}/N$, such that the read cost $R=O(1)$ and hence the last level false positive rate can be set to one. Plugging in $p_{i}$ with the optimized allocation in Equation~\ref{eq:p_i(p_L)} into Equation~\ref{eq:new total filter memory}, we can obtain a closed form solution by taking the dominating terms in the summation of $\sum_{i=1}^{L} ln(p_{i}) \frac{c^{(L-i)(L-i-1)/2}}{T^{L-i}}$ such that 
\begin{equation*}
 \frac{M_{filter}}{N} \approx \frac{-1}{ln(2)^2} (\frac{1}{T}ln(\frac{1}{T}p_{l}) + \frac{c}{T^2}ln(\frac{c}{T^2}p_{l})
\end{equation*}
Setting $p_{l}$ to one and the partial derivative to zero, we can find the above function has a maxima at 1.52 bits per entry. With the bloom filter optimization, the required memory budget to achieve $O(1)$ zero-look up complexity is very affordable. In industry, a standard LSM-tree key-value storage engine typically uses 10 bits per data entry such that the FRPs are all lower than one. When the memory budget is constrained, such as on mobile or IoT devices, the read point query becomes much faster with the bloom filter optimization compared to allocating a fixed number of bits per entry uniformly across all levels.

\smartparagraph{CPU Optimization.} Each run is associated with a bloom filter stored in the main memory (DRAM) to reduce disk I/O. Researchers have observed that the improvement in computational power has slowed down~\cite{theis2017end}, and, as a result, the filter CPU costs may become a new bottleneck in the future~\cite{dayan2021chucky,dayan2021end, zhu2021reducing}. Instead of designing new filters to minimize the CPU cost, Autumn directly reduces the CPU cost by storing fewer bloom filters in memory, as the number of runs decreases from $O(logN)$ to $O(\sqrt{logN})$ (recall there is one run per level). Hence, while the worst case point read complexity in terms of the numbers of disk I/Os are the same between Leveling and \policy, \policy has a lower CPU cost and performs better than Leveling which is also validated in the experiments.

\smartparagraph{Range Reads Cost.} Autumn focuses on optimizing both point reads and range reads. Range reads can be classified into two categories: i) Short Range Reads and ii) Long Range Reads. A short range reads implies data entries within the target key range are, on average, stored in one data block on disk for each run. Hence, each run has one potential data block that needs to be examined for short range reads. As a result, \policy improves the short range reads complexity from $O(logN)$ in Leveling to $O(\sqrt{logN})$ since the number of levels with \policy is less than that of Leveling.

A long range reads implies that data entries within the target key range are, on average, stored in multiple data blocks for each run. Hence, the long range reads have two sources of disk I/Os from each run. One involves seeking to the initial position, which is one disk I/O, and then reading consecutive blocks till the ending boundary is reached. While the number of consecutive blocks depends on the uses' query, \policy reduces the number of seeks that need to be performed.

\smartparagraph{Write Amplification.} Recall that the write amplification is the average number of disk writes for each data entry in the LSM-tree. We observe that the data entries in level $L$ need to be compacted on average $T$ times since level $L$ is $T$ times larger than level $L-1$. Similarly, level $L_{L-i}$ needs to be compacted on average $T/c^i$ times. As a result, the number of merge operations in level 1 dominates the write amplification, and the overall number of compaction per data entry is, therefore, $O(T/c^{\sqrt{log_{1/c}(\frac{N}{B \cdot T})}})$.

\smartparagraph{Analyzing Write Amplification Complexity.} Write Amplification is crucial for scaling the storage engine and in this subsection we demonstrate why \policy is scalable for write. An initial glance at $O(T/c^{\sqrt{log_{1/c}(\frac{N}{B \cdot T})}})$, the $O(1/c^{\sqrt{log_{1/c}(N/\frac{N}{B \cdot T})}})$ term may seem large. However, if we closely examine this complexity bound, then we can observe that $O(1/c^{\sqrt{log_{1/c}(\frac{N}{B \cdot T})}})$ is actually sub-linear to $\frac{N}{B \cdot T}$. We observe that $O((\frac{N}{B \cdot T})^{1/x})$ for any constant $x$ is equivalent to $O(1/c^{\frac{1}{x}log_{1/c}(\frac{N}{B \cdot T})})$. As an example, $O(\sqrt{\frac{N}{B \cdot T}})$ is equivalent to $O(1/c^{\frac{1}{2}log_{1/c}(\frac{N}{B \cdot T})})$. Since these complexities share the same base $(1/c)$, we can simply compare the exponents. \policy's $O(T/c^{\sqrt{log_{1/c}(\frac{N}{B \cdot T})}})$ exponent with respect to the base $1/c$ is $\sqrt{log_{1/c}(\frac{N}{B \cdot T})}$, while for $O((\frac{N}{B \cdot T})^{1/x})$, its exponent with respect to the base $1/c$ is $\frac{1}{x}log_{1/c}(\frac{N}{B \cdot T})$. It is clear that the complexity of $O(T/c^{\sqrt{log_{1/c}(\frac{N}{B \cdot T})}})$ is sublinear and it is less than $O((\frac{N}{B \cdot T})^{1/x})$ for any constant $x$. Therefore, we conclude that Autumn with our proposed \policy policy is scalable for updates.

\smartparagraph{Autumn Implicitly Prioritizes Compaction in Lower Levels.}
As the volume of data entries grows and the levels in Autumn expand, the capacity of each level increases accordingly. Notably, the capacities of larger levels expand at a faster rate than those of lower levels. This disparity in growth rates leads to less frequent compaction triggers for larger levels, while lower levels undergo compaction more frequently. Although this might cause higher write amplifications, it can enhance overall write throughput. Given the fixed memory budget and the need for LSM-tree-based key-value stores to maintain write-ahead logs for unflushed memory tables, these systems often implement a rate limiter. When an excessive number of files remain unflushed to disk as a sorted run, the rate limiter is activated, either restricting the write rate or pausing all writes to allow for compactions in lower levels, to avoid potential data loss. Previous research has demonstrated that prioritizing flushes can significantly enhance write throughput~\cite{balmau2019silk}. Hence, by implicitly having more compactions in the lower levels, Autumn effectively reduces write stalls and boosts write throughput. We will further explore this benefit in the YCSB macro experiment section.

\subsection{Further Improvements in Performance} 
\label{sec:pin}


Autumn can further improve write costs by keeping multiple sorted runs in level 0. When the immutable memtable flushes to level 0, no internal compaction (sort-merge) within level 0 will be triggered. The primary purpose of this flush is to ensure data durability. With this approach, there are multiple sorted runs at the first level. This is in fact how LevelDB and RocksDB are implemented, where they keep a constant number of sorted runs in level 0 to speed up the flushing from memory to disk. Since the number of sorted runs in level 0 is constant, this does not affect the read complexities.

Autumn benefits from this design choice. The main observation is that Autumn distributes more compactions in lower levels, and the first level of Autumn frequently gets compacted as the data size increases, and the compaction costs in the first level dominate the overall write costs. For instance, assuming \policy with $c=0.8$, $T=2$, and six levels, the first level consists of $0.18\%$ of total data, and the write amplification from the first level consists of $30\%$ of the total writes. In contrast, the traditional merge policies, such as Leveling and Tiering, spend the same amount of write amplification at each level, and the write costs are distributed uniformly across all levels. 
As a result, having multiple sorted runs in level 0 further improves the overall performance of Autumn with \policy policy, by reducing the total write amplification as the flush operation does not perform any compactions (the first level essentially becomes tiered). Moreover, LSM-tree storage engines like RocksDB offer the option to pin Level 0 metadata in the block cache. Activating this feature can accelerate file reading from Level 0. Given that the capacity of Level 0 is relatively small compared to the total storage size, pinning its metadata in main memory occupies a minimal amount of the block cache's capacity.

Prior research~\cite{im2020pink, yao2020matrixkv} proposed pinning the first few levels in non-volatile memory to improve write performance. These works pin a significant portion of data stored in non-volatile memory (NVM). For instance, Pink~\cite{im2020pink} pins the first three levels of an LSM-tree in NVM and also stores hundreds of megabytes in DRAM for level lists which help locating the meta data blocks. In contrast, LevelDB, RocksDB, and Autumn have at most 256 MB (determined by the $max\_bytes\_for\_level\_base$ knob) for the files in the first LSM-tree level. The experimental section describes the implementation details in which we modified the forked codebase from LevelDB and RocksDB (Section~\ref{sec:eval}). As a result, we do not compare Autumn the works incorporating specialized hardware, but we believe these designs might also benefit from the uneven distribution of compactions as lower levels stored in non-volatile memory have high write bandwidth.

%% file: experiments.tex
\section{Experiments} \label{sec:eval}
In this section, we present experimental results and showcase that Autumn is optimized for reads and remains scalable for space amplifications and writes.

\subsection{Implementation Details}
We build Autumn on top of RocksDB~\cite{matsunobu2020myrocks} (forking in October 11, 2023) and LevelDB~\cite{levelDB} (forking the most recent 1.23 version). To incorporate how \policy modifies the level capacity as the stored data increases, Autumn modifies the \textit{MaxBytesForLevel} function, stores the \textit{max\_level\_in\_use\_}, and attaches the level information on each file to calculate the level capacity according to Equation~\ref{eq:new capacity}. \policy has the same capacity ratio as Leveling when $c$ is set to 1. In addition, to observe the full effect of changing the capacity ratio across levels, we disabled the seek compaction optimization, which is by default disabled in RocksDB due to its complexity in the critical code path~\footnote{https://github.com/facebook/rocksdb/blob/main/HISTORY.md}. The main reason that we also implemented Autumn on LevelDB is to perform a fair comparison on point reads with Monkey~\cite{dayan2017monkey} which introduced the bloom filter optimization and Monkey is implemented on top of LevelDB. Both RocksDB and LevelDB, by default, do not have internal compaction for the first level, and also, the compaction trigger for the first level to the second level is not by capacity but rather by the number of runs stored. Moreover, there is a rate limiter to limit the number of sorted runs. For example, the \textit{max\_write\_buffer\_number} knob limits the number of memetables in main memory and the \textit{level0\_stop\_writes\_trigger} bounds the number of runs in level 0.


\smartparagraph{Configurations}
To ensure fair comparisons across all experiments, every storage instance will maintain identical configurations with the exception of the $c$ parameter. This parameter distinguishes our novel capacity ratio assignment from existing methods. Additionally, we have disabled block caches and compression, and have turned off the dynamic leveling approach in RocksDB, which aims to minimize space amplification. We have enabled direct IO for RocksDB, which bypasses the OS page cache; however, since LevelDB does not support this feature, it utilizes buffered IO. When $c$ is set to 1.0, the experiments are conducted directly on the storage engine compiled with the RocksDB source code, as of October 11, 2023.

\subsection{Micro Benchmarks}
To compare and contrast Autumn with RocksDB, we first conduct small-scale micro benchmarks with the default benchmark tool, \textit{db\_bench}. For \textit{db\_bench}, we disabled the compression to better understand and analyze the performance. In addition, we set the options in the storage engine to $OptimizeForSmallDb$ configuration (recommended by RocksDB) which sets the target file size to 2MB and the maximum base capacity to 10 MB.

In addition, we use six different operations in the micro benchmark:
\begin{itemize}
    \item \textit{FillSeq}: Write data entries in sequential order.
    \item \textit{FillRandom}: Write data entries in random order.
    \item \textit{ReadRandom}: Perform random point queries.
    \item \textit{SeekRandom}: Perform random seeks.
    \item \textit{SeekRandomNext10}: Perform random small range queries.
    \item \textit{SeekRandomNext100}: Perform random long range queries.
\end{itemize}
FillSeq and FillRandom write to two independent database instances. The read operations are performed on the database instance with data entries written from FillRandom. For \textit{SeekRandomNextN}, the iterator first randomly seeks to a desired key, and then the iterator performs the next operation $N$ times while the iterator is valid. 

\begin{figure*}[h]
\centering
    \subfloat[Write Performance]{ \includegraphics[width=0.32\textwidth]{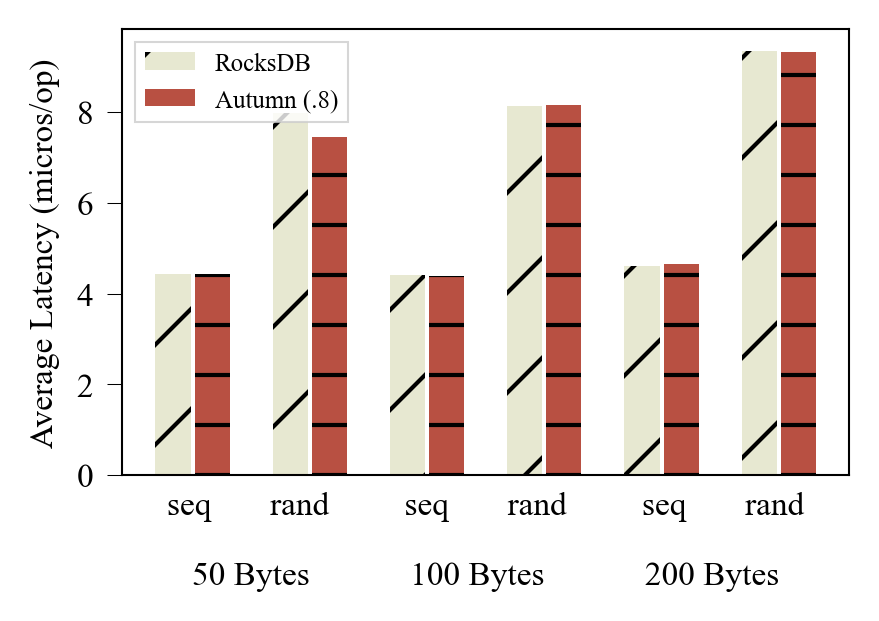}}
    \hfill
    \subfloat[Point Query]{ \includegraphics[width=0.32\textwidth]{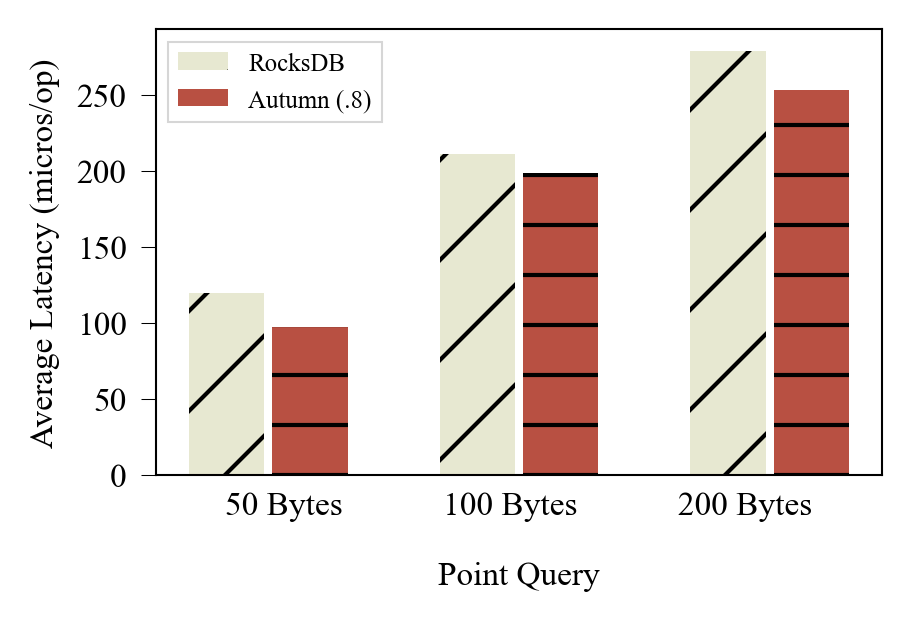}}%
    \hfill
    \subfloat[Range Query]{ \includegraphics[width=0.32\textwidth]{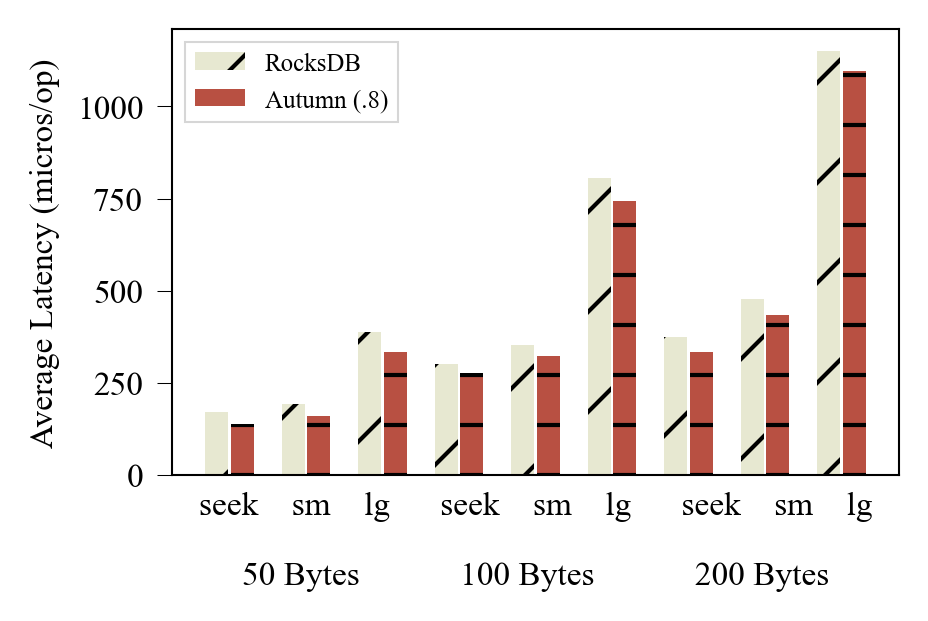}}%
    \hfill
\caption{The evaluation results between RocksDB and Autumn ($c=.8$) with six different operations using db\_bench micro benchmark.}
\label{exp:microbench}
\end{figure*}

\smartparagraph{Micro Benchmarks Running Environment.} The micro bench experiments are run on an AWS i3en.large ec2 instance with Ubuntu 22.04 operating system, 2.5 GHz Intel Xeon Platinum 8259CL CPU and 1.14 TiB Amazon EC2 NVMe Instance Storage.

\subsubsection{Different operations with varying key-value sizes}
We set the default policy parameters $T = 2$ for both Autumn and RocksDB and use $c=.8$. We vary the value size among 50 bytes(Zippy, UP2X), 100 bytes(UDB, VAR), and 200 bytes(APP, ETC), as these value sizes reflect the average value size of different data workloads in Facebook’s production systems~\cite{atikoglu2012workload, cao2020characterizing}. Recall the storage options are set to $OptimizeForSmallDb$ configurations and we disabled data compression and block cache, consistent with the macro benchmarks below. To understand the worst-case performance of point queries, we omit the bloom filters in this micro benchmark, and in later sections, we showcase the benefit of the bloom optimization (Section~\ref{sec:autumn}) in large-scale experiments. 

\smartparagraph{Write Performance.} The write performance is measured from two operations, FillSeq and FillRandom, each containing two million data entries in sequential and uniformly random order. We use average latency per operation as the metric and lower latency indicates better performance. As shown in Figure~\ref{exp:microbench} (a), Autumn and RocksDB have lower latency when writing sequentially than uniformly random. They also share the similarity of having higher write latency as the value size increases, in which more bytes need to be written. In general, Autumn and RocksDB share similar latency among different value sizes and write patterns. 



\smartparagraph{Point Query.} The point query performance is measured from ReadRandom operations which read over one million random keys. We use average latency per operation as the metric, and lower latency is more desirable. Also, both Autumn and RocksDB do not have bloom filters attached. As shown in Figure~\ref{exp:microbench} (b), Autumn performance is constantly better than RocksDB for all value sizes. For instance, with the value size of 50 bytes, Autumn's random read average latency is 97.526 microseconds per operation, and RocksDB's random read average latency is 119.937 microseconds per operation, which is about 19\% improvements. When runs have no bloom filter, the point read latency becomes proportional to the number of levels. Since Autumn, by design, reduces the total number of levels, Autumn outperforms RocksDB. 

\smartparagraph{Range Query.} The range query performance is measured for three operations, namely: Seek, SeekAndNext10, and SeekAndNext100. These operations are performed one million times. Similarly to previous metrics, we use average latency per operation to measure performance, and the lower y-axis is better. As shown in Figure~\ref{exp:microbench} (c), Autumn performance is better than RocksDB across all value sizes and for all three operations. Among the three operations, the seek operation has the best performance improvement. For example, with a value size of 50 bytes, Autumn's seek latency is 138.864 microsecond per operations and RocksDB's seek latency is 170.742 microsecond per operations, an improvement of by about 19\%. As the number of next operations increases, the improvement slightly decreases, which is due to more data need to be fetched. Moreover, as the value sizes increase, the improvement also decreases. Larger key value sizes lead to more number of data blocks need to be fetched to find all the target data entries. For example, when the value size is set to 200 bytes, small range reads improve by 9\%, and long range reads improve by 5\%.

\begin{figure}[h]
\centering
    \subfloat[FillRandom]{ \includegraphics[width=0.23\textwidth]{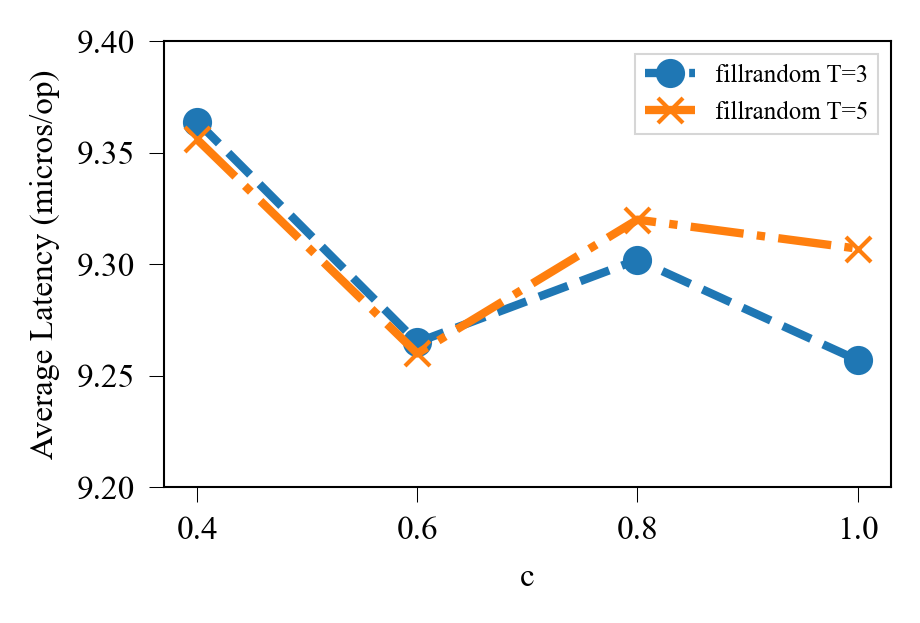}}%
    \hfill
    \subfloat[Small Range Reads]{ \includegraphics[width=0.23\textwidth]{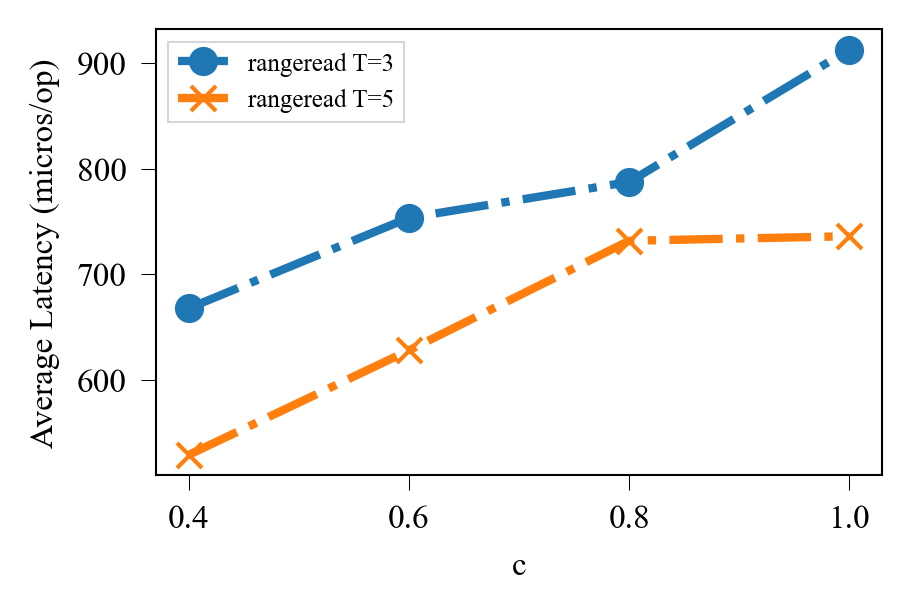}}%
        
\caption{The effect of $c$ and $T$ on small range query.}
\label{exp:vary_c_k}
\end{figure}

\subsubsection{Write and Read Sensitivity to $c$ and $T$}
There are two parameters $c$ and $T$ in the \policy policy. Parameter $c$ controls how gradually the small levels expand as more data are stored, and $T$ determines the capacity ratio between the last two levels. In this section, we explore the effect of varying $c$ or $T$ while holding the other parameter constant on random writes and small range reads. For this analysis, we first use FillRandom to write two GB of key-value pairs, each with the db\_bench default value size of 100 bytes. In Figure~\ref{exp:vary_c_k}, the x-axis is $c$ which varies from 0.4 to 1.0, and the y-axis is the average latency per operation. Lower average latency indicates better performance. Figure~\ref{exp:vary_c_k} (a) depicts two lines of random writes' performance with either $T=3$ or $T=5$ and similarly, Figure~\ref{exp:vary_c_k} (b) depicts two lines of small range reads' performance with either $T=3$ or $T=5$. 


We can observe that in general, when $T$ is fixed, lower $c$ gives better read performance and worse write performance. Recall that the number of levels in \policy is $O(\sqrt{-log_{c}(\frac{N}{B\cdot k})})$ and small range query performance is proportional to the number of levels. As expected, when $c$ decrease, the number of levels also decreases; hence small range query performance improves. In addition, when $T$ increases, the number of levels decreases, and therefore small range query performance improves. 


\subsection{Macro Benchmark}
In this section, we compare Autumn and RocksDB performance with large-scale data using the YCSB benchmark~\cite{cooper2010benchmarking} (Yahoo! Cloud Serving Benchmark), which is a widely used macro-benchmark suite to study key-value store performance. This macro benchmark further compares the performance between Autumn and RocksDB on real-world workload traffic.

\smartparagraph{Macro Benchmarks Running Environment.} The macro benchmark experiments are run in the same environment as the micro benchmarks above.

\begin{figure}[tbph]
\centering
\includegraphics[width=0.45\textwidth]{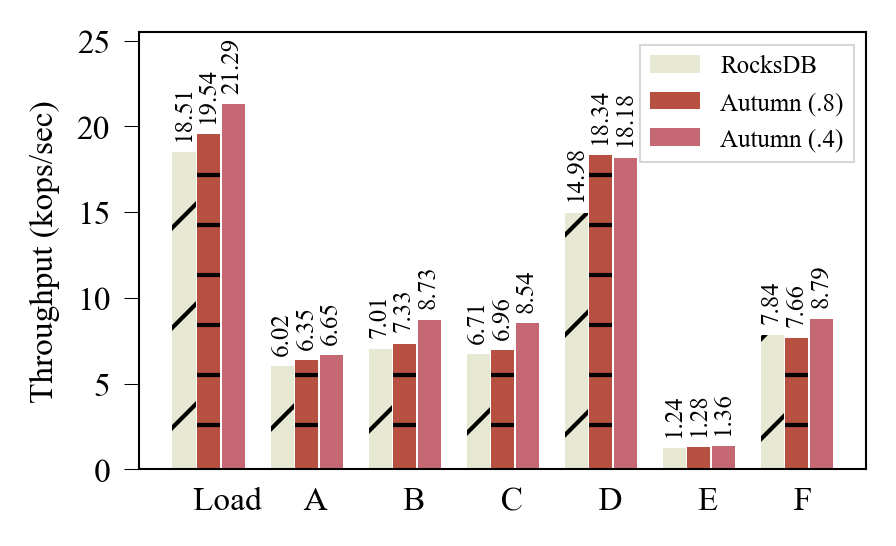}
\caption{YCSB Workloads Performance. This figure shows the performance comparison between Autumn and RocksDB on various YCSB core workloads. The x-axis presents the different workload names, and the y-axis shows the throughput using a base unit of a thousand ops per second (kops/sec). The actual throughput (kops/sec) achieved is placed on top of each bar.}
\label{exp:ycsb_default}
\end{figure}

\smartparagraph{YCSB Core Workloads A-F.} To further compare Autumn with RocksDB, we conduct experiments using YCSB default settings and workloads. We set $T=5$ for Autumn and RocksDB, and we disabled data compression and block cache for a better understanding of the performance. To better illustrate Autumn's capability for read improvements, we set $c=0.8$ and $c=0.4$ for Autumn. We expect read performance to improve as c decreases. We first write a 80 GB dataset with 1KB values (the default value sizes) for loading and then evaluate workloads A-F with one million operations. The characteristics of different workloads are:
\begin{itemize}
    \item \textit{Load}: Write 80 GB of data entries into the corresponding key-value store, and the data entry size is 1KB, 24 bytes key and ten fields each with 100 bytes.
    \item \textit{A}: Workload A is update heavy. 50\% point reads and 50\% updates. Keys follow Zipfian~\cite{zipf2016human} distribution.
    \item \textit{B}: Workload B is read mostly. 95\% point reads and 5\% updates and keys follow Zipfian distribution.
    \item \textit{C}: Workload C is read only. 100\% point reads with Zipfian distribution.
    \item \textit{D}: Workload D is read latest. 95\% point reads and 5\% insertions. Requests are temporally weighted.
    \item \textit{E}: Workload E is range reads. 95\% random range queries and 5\% insertions. On average about 100 keys are read for each range read.
    \item \textit{F}: Workload F is read-modify-write workload and it has 50\% reads and 50\% read-modify-writes.
\end{itemize}

As shown in Figure~\ref{exp:ycsb_default}, our first observations indicate that Autumn consistently performs at par or better than RocksDB across all default YCSB workloads. Specifically, during the load phase, Autumn achieves a significantly higher throughput than RocksDB, with a 15\% increase in write throughput when comparing Autumn with $c=0.4$ to RocksDB. Upon examining the database statistics, it is evident that Autumn with $c=0.4$ experiences 1265 write stalls, while Autumn with $c=0.8$ encounters 2592 write stalls, and RocksDB suffers from 3403 write stalls. The primary causes of these write stalls are linked to an excessive number of memtables and pending compaction bytes, which are bytes scheduled for near-future compaction. These factors contribute to delays in processing write operations, thereby affecting the overall performance. By strategically prioritizing compactions at lower levels and incrementally expanding the capacity of each level to buffer more entries before compaction, Autumn surprisingly achieves an increase in write throughput. Also, the write stalls are unlikely to be triggered with small data load and hence we didn't observe this phenomena in the micro benchmark experiments.

On the read-only workload (Workload C), we do find Autumn achieves less improvements than we had expected. Looking into the level summaries, we find that RocksDB has 5 levels, Autumn (.8) has 4 levels, and Autumn (.4) has almost 3 levels in which there is a small file of 64.3 MB residing in level 4. In our analysis, read performance should be proportional to the number of levels. However, we can observe that Autumn with $c=.8$ has only a small improvement on the overall throughput compared to RocksDB (4\%) and Autumn with $c=.4$ achieves a 24\% increase in the overall throughput. This observation can be attributed to the differences between the settings used for the worst-case analysis and the YCSB workload. In the worst-case scenario, the worst-case read performance occurs when keys are not found in the key-value store, necessitating searches across all levels. Conversely, in the YCSB workload which simulates the skewed distribution in real-wrold traffics, almost all keys are presented in the key-value store. Therefore, once a key is located, no additional disk IOs are required, minimizing the improvements between Autumn and RocksDB.

In Workload E, which focuses on range reads, we also observed an improvement that is not as substantial as anticipated. Two primary factors contribute to this modest enhancement. First, nearly all keys involved in range reads are found in the database, contrary to the worst-case scenario analysis which assumes that target key ranges may not be stored. Additionally, in YCSB, data entries are set at 1KB, with a disk block size of 4KB, and a scan length of 100, meaning that on average, 25 disk blocks are required. Consequently, the number of disk I/O operations is largely influenced by the invocation of 'next' operations on the range read iterator, rather than the seek operation. The larger size of data entries inherently raises the costs of data retrieval, which results in lower throughput for both Autumn and RocksDB when compared to their point read performances.

Furthermore, in Workload D, which involves reading recently written keys, Autumn demonstrates improvements exceeding 20\%. This enhancement is largely due to the temporal locality in writes and reads for this workload. Since Autumn strategically increases the capacity of each level, Autumn with $c=0.4$ achieves 40\% hits at Level 1, while Autumn with $c=0.8$ secures 33\% hits at the same level, whereas RocksDB only has 9\% hits in Level 1. Because point read queries halt immediately upon locating the key, this results in a notable increase in throughput for Autumn.

We also analyzed the space amplification of these three key-value stores following the load phase. Although the expected logical storage size was 80 GB, the physical storage measurements were slightly higher: RocksDB stored 87.42 GB, Autumn with $c=.8$ used 87.49 GB, and Autumn with $c=.4$ used 87.61 GB. While Autumn exhibits a slight increase in space amplification, the overall differences among the systems remain minimal. RocksDB, a production-level storage engine deployed in numerous industry databases~\cite{zhou2021foundationdb, feng2015benchmarking, taft2020cockroachdb}, sets a high standard; our design reliably matches or even exceeds its performance consistently across various workloads.

\begin{figure*}[hbtp]
\centering
    \subfloat[Write Performance]{ \includegraphics[width=0.32\textwidth]{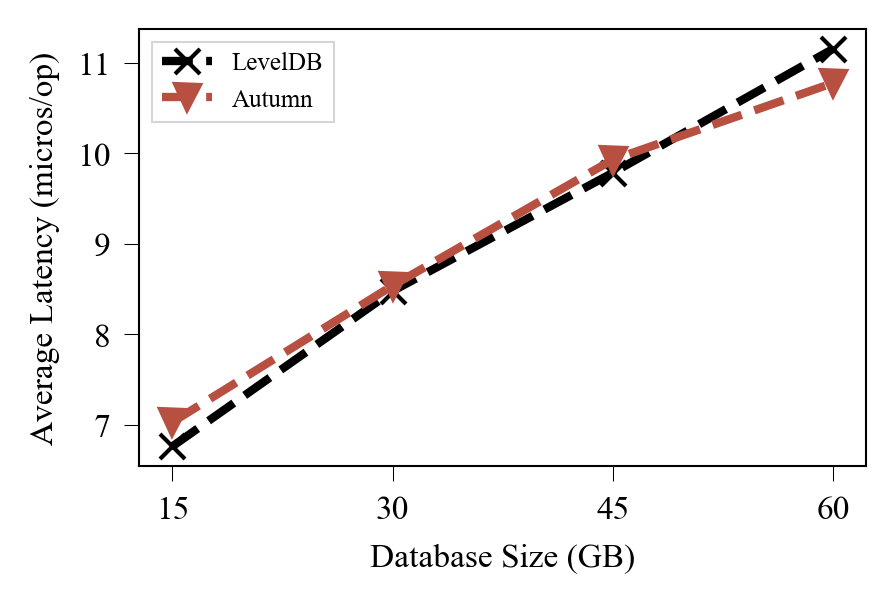}}
    \hfill
    \subfloat[Point Query]{ \includegraphics[width=0.32\textwidth]{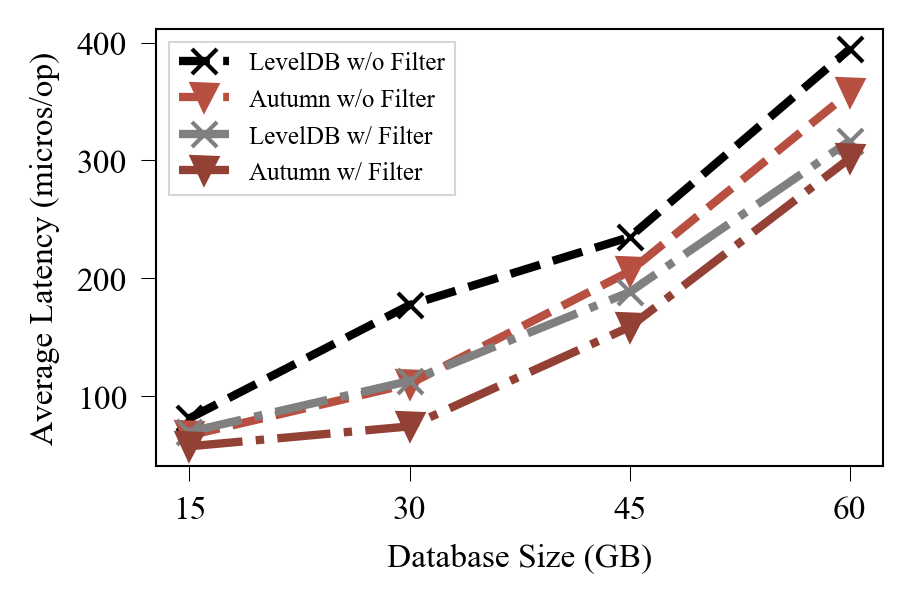}}%
    \hfill
    \subfloat[Small Range Query]{ \includegraphics[width=0.32\textwidth]{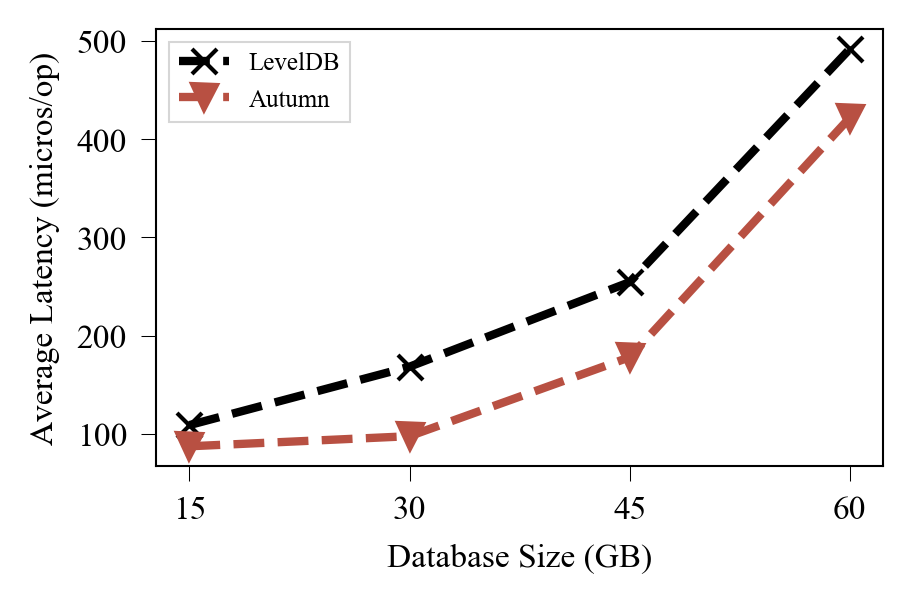}}%
    \hfill
\caption{Performance on \textit{db\_bench} Macro Benchmarks with Varying Database Size and Different Operations. The Performance is Measured in Average Latency.}
\label{exp:macrobench}
\end{figure*}

\begin{table}[htbp]
    \begin{center}
    \begin{adjustbox}{width=0.38\textwidth}
    \begin{tabular}{ |l|c|c|c| } 
    \hline
    Latency ($\mu$s/op) & Avg. & 95\% & 99\% \\
    \hline
    \multicolumn{4}{|c|}{Point Read Latency in Workload A} \\ \hline
    RocksDB & 157 & 513 & 640 \\ \hline
    Autumn .8& 148 & 475 & 612 \\ \hline
    Aurumn .4 & 140 & 443 & 536 \\ \hline
    \multicolumn{4}{|c|}{Point Read Latency in Workload C} \\ \hline
    RocksDB & 149 & 488 & 595 \\ \hline
    Autumn .8& 143 & 454 & 579 \\ \hline
    Aurumn .4 & 117 & 332 & 459 \\ \hline
    \multicolumn{4}{|c|}{Range Read Latency in Workload E} \\\hline
    RocksDB & 846 & 1907 & 2229 \\ \hline
    Autumn .8& 822 & 1851 & 2141\\ \hline
    Autumn .4& 773 & 1759 & 2006 \\
    \bottomrule
    \end{tabular}
    \end{adjustbox}
    \end{center}
    \caption{Average and Tail Latency}
    \label{tab:Tail Latency}
\end{table}

\smartparagraph{YCSB Average and Tail Latency.} LSM-tree based key-value stores deployed in production environments should provide high-quality services to their customers. Average and tail latency for point and range queries are crucial statistics in delivering good service, and tail latency can be crucial for latency-critical applications. Lower latency is more desirable for user experience. In Table~\ref{tab:Tail Latency}, we listed the average and tail latency numbers (microsecond per operation) for point read or range read operations in Workload A, C, and E. These workloads demonstrate a representative set of different operations, ranging from Workload A, which is update heavy, Workload C, which is read-only, and Workload E, which has range reads with some insertions. We find that Autumn achieves lower latency numbers across all reported quantile values and among these different workloads. For instance, comparing RocksDB with Autumn ($c=.4$), RocksDB 99\% tail latency is 1.2x, 1.3x, and 1.1x worse for Workload A, C, and E, respectively. 
These improvements can positively affect the quality of user experience.




\subsection{Bloom Filter Optimization}
In this section, we compare Autumn, which is forked from LevelDB, to its original counterpart. Monkey~\cite{dayan2017monkey}, which originally proposed the bloom filter optimizations, is implemented in LevelDB. It is important to note that LevelDB does not offer the option to enable direct IO; therefore, buffered IO is utilized. As demonstrated in Section 3, Autumn is expected to achieve performance comparable to LevelDB when bloom filter optimizations are employed. In the following experiments $T=2$ is used for both LevelDB and Autumn.

\textbf{LevelDB \textit{db\_bench} benchmark.} Similar to previous benchmarking steps, Autumn and LevelDB each create four data storage instances with 15, 30, 45, and 60 GB using \textit{Fillrandom} operations (keys are uniformly random). The key is 16 bytes, and the value is 100 bytes. After all data entries are written, we collect the average latency for writes, point reads with and without bloom filters (setting the number of bits per data entry to two and applying bloom filter optimization), and small range reads. As shown in Figure~\ref{exp:macrobench}, the x-axis is the database size, and the y-axis is the average latency measured in microseconds per operation in all sub-figures. Lower average latency indicates better performance. Figure~\ref{exp:macrobench} (a) shows the write performance, and Autumn has comparable write performance compared to LevelDB when buffer IO is employed. For instance, Autumn has a tiny performance improvement compared to LevelDB at 60 GB, in which the average latency decreased from LevelDB's 11.153 microseconds per operation to Autumn's 10.776 microseconds per operation. Similar to RocksDB, LevelDB also activates rate limiters when too many memtables are not flushed onto disk. The write stalls introduced by rate limiters are used to ensure data become durable on disk and avoid potential data loses in the case of machine failures. With a large amount of data, Autumn gradually mitigates these write stalls by implicitly prioritizing compactions in the lower levels. As shown in Figure~\ref{exp:macrobench} (b), Autumn outperforms LevelDB both when bloom filters are not used and when bloom filters (with bloom filter optimization) are applied to each sorted run. When bloom filters are attached, Autumn's point read performance improvement decreases while Autumn still exceeds LevelDB in performance. This behavior is expected since Autumn has a faster convergence when summing the false positive rates, and with more data and more levels, their performance eventually converges. Moreover, we also observe a steady performance improvement for range read, shown in Figure~\ref{exp:macrobench} (c), which compares the small range query using \textit{SeekAndNext10} operations. When the database size is 30 GB, Autumn has a 42\% improvement on small range query compared to levelDB. These experiments showcase that, with a large amount of data, Autumn offers fast point and range read speeds while maintaining low write latencies.



%% file: relatedWorks.tex
\section{Related Works} \label{sec:relatedWorks}
Many existing research works have explored other opportunities to improve LSM-trees performance. In this section, we discuss related works in this domain, and many are, in fact, orthogonal compared to Autumn, and Autumn may incorporate these approaches to further improve the performance.

\textbf{Reduce Write Amplification.} Researchers have primarily focused on optimizing the write cost of the LSM-tree key-value store. WiscKey~\cite{lu2017wisckey} separates values from keys by storing keys in an append-only log. As a result, WiscKey reduces the write amplification, and because of the fast random reads property in modern SSDs~\cite{chen2011essential}, read query performance remains robust (small regressions). Other works~\cite{chan2018hashkv, xanthakis2021parallax} also adopt this design to use an external log for storing values. Autumn may adopt the key value separation approach to further improve write performance. In addition, existing research works, such as MatrixKV~\cite{yao2020matrixkv}, Pink~\cite{im2020pink}, and NoveLSM~\cite{kannan2018redesigning}, explored opportunities in the hybrid DRAM-NVM-SSD tired storage systems to store non-trivial portion of the data in NVM to reduce the write cost. Autumn's \policy merge policy may bring even better benefits in such settings, as Autumn distributes higher write costs among smaller levels. When smaller levels with higher write costs can be pinned in a device with stronger processing power, the overall write costs significantly decreases.

\textbf{Reduce Write Stall.} A line of work aims to reduce the impact of write stall when serving clients' requests. SILK~\cite{balmau2019silk} mitigates write stalls by postponing flushes and compactions to periods of low load and allowing smaller levels compactions to preempt higher levels compactions. Kvell~\cite{lepers2019kvell} reduces write stalls by minimizing the CPU computation cost, in which it stores unsorted key-value pairs on disks to reduce CPU cost. As a result, it sacrifices the read performance.

\textbf{Reduce Space Amplification} RocksDB~\cite{dong2017optimizing} proposed to empirically reduce storage space amplification in LSM Trees by bounding the capacity of smaller levels based on the size of the last level. Since the approximate size of the last level is stored in the metadata, both LevelDB and Autumn can embrace this optimization with some modifications. 

\textbf{Data Sketch for LSM-trees.} Data Sketches provide approximate answers to a wide range of question with theoretically proven error bounds~\cite{zhao2021kll, zhao2021spacesaving, liu2023hypercalm, liutreesensing, karnin2016optimal, metwally2005efficient,cormode2005improved, zhao2022differentially}. For instance, membership queries can be solved using bloom filters~\cite{bloom1970space}. Filters are now widely used in LSM-trees based key-value stores to improve point read performance. New point filters, such as cuckoo filters~\cite{fan2014cuckoo} and quotient filters~\cite{bender2011don}, are developed to improve the accuracy level given a fixed memory budget. To improve range read performance, researchers have designed range filters, pioneered by ARF~\cite{alexiou2013adaptive} and later followed by SuRF~\cite{zhang2018surf} and Rosetta~\cite{luo2020rosetta}, to skip disk access if the storage or file do not contain the target key range. However, range filters require much more space budget compared to point filters to support arbitrary key ranges. Given a fixed memory budget, range filters sacrifice the point filter query accuracy.

\textbf{Hardware Acceleration and Workload Optimization.} Key-value stores can utilize new advancements in specialized hardware to speed up their performance, such as FPGAs~\cite{huang2019x, jiang2023data}, smart network interface cards~\cite{li2017kv}, GPUs~\cite{ashkiani2018gpu}, persistent memory~\cite{vogel2022plush} and heterogeneous storage~\cite{yoon2018mutant}. Moreover, techniques on multi-core CPUs~\cite{zhang2014top} and lock-free synchronization~\cite{golan2015scaling} can also improve LSM-trees performance. Furthermore, when reads are skewed, Zhang et al.\ ~\cite{zhang2022bi} propose to improve read performance by moving hot data entries back to smaller levels. These hardware acceleration and workload dependent optimization approaches are orthogonal to Autumn, as the \policy merge policy can build on top of these different underlying hardware architectures.

In the future, we would also like to investigate how \policy improves the performance of LSM-trees in multi-tier storage (e.g., DRAM, NVM, SSD) or helps reduce the costs in streaming database~\footnote{https://www.risingwave.dev/docs/current/data-persistence/} (e.g., data stored in local drive, distributed file system, and Amazon simple storage). Since Autumn automatically introduces more compactions in the lower levels and the lower levels with less amount of data usually reside in faster devices, we believe that Autumn may also lead to better performances in such systems.


%% file: conclusion.tex
\section{Conclusion} \label{sec:conclusion}
In this paper, we introduce Autumn, a novel key-value store featuring the innovative \policy compaction policy that redefines data organization within LSM-Trees to enhance the worst-case complexity for both point and range reads. 
Additionally, the introduction of the delayed compaction and the fact that the \policy policy prioritizes compactions at the smaller levels, significantly boosting Autumn's efficiency in read and write performances. 
We conducted thorough evaluations using the \textit{db\_bench} and \textit{YCSB} benchmarks to benchmark Autumn against Google's LevelDB and Meta's RocksDB. 
Our findings demonstrate that Autumn not only optimizes read operations but also maintains robust write performance. Consequently, the \policy compaction policy equips Autumn to substantially enhance key-value storage performance, making it well-suited for OLTP and HTAP workloads.